\documentclass[manuscript]{acmart} 

\AtBeginDocument{%
  \providecommand\BibTeX{{%
    \normalfont B\kern-0.5em{\scshape i\kern-0.25em b}\kern-0.8em\TeX}}}

\setcopyright{acmlicensed}
\copyrightyear{2022}
\acmYear{2022}
\acmDOI{10.1145/3523227.3551474}

\acmConference[RecSys '22]{Sixteenth ACM Conference on Recommender Systems}{September 18--23, 2022}{Seattle, WA, USA}
\acmPrice{15.00}
\acmISBN{978-1-4503-XXXX-X/18/06}

\usepackage{bm}
\usepackage{acronym}
\begin{document}

\title[Discovery Dynamics]{Discovery Dynamics: Leveraging Repeated Exposure for User and Music Characterization}

\author{Bruno Sguerra}
\authornote{Contact author: \href{research@deezer.com}{research@deezer.com}}
\affiliation{
  \institution{Deezer Research}
    \city{}
  \country{France}
}

\author{Viet-Anh Tran}
\affiliation{
  \institution{Deezer Research}
    \city{}
  \country{France}
}

\author{Romain Hennequin}
\affiliation{
  \institution{Deezer Research}
    \city{}
  \country{France}
}

\renewcommand{\shortauthors}{B. Sguerra et al.}


\begin{abstract}
Repetition in music consumption is a common phenomenon. It is notably more frequent when compared to the consumption of other media, such as books and movies. In this paper, we show that one particularly interesting repetitive behavior arises when users are consuming new items. Users’ interest tends to rise with the first repetitions and attains a peak after which interest will decrease with subsequent exposures, resulting in an inverted-U shape. This behavior, which has been extensively studied in psychology, is called the mere exposure effect. In this paper, we show how a number of factors, both content and user-based, well documented in the literature on the mere exposure effect, modulate the magnitude of the effect. Due to the vast availability of data of users discovering new songs everyday in music streaming platforms, these findings enable new ways to characterize both the music, users and their relationships. Ultimately, it opens up the possibility of developing new recommender systems paradigms based on these characterizations. 

\end{abstract}


\maketitle

\section{Introduction}
One fundamental difference between the consumption of music and other media, such as movies, books, and podcasts, is repetition. While it is rare for moviegoers and book readers to revisit the same piece more than a couple of times, listening repetitively to music, not only does not seem to bother consumers much but is a rather common phenomenon. To illustrate, the term ``spoiler'', i.e., beforehand information about the plot, is commonly employed in the context of consumption of movies, series, and books. As the term implies, spoilers have the potential to ruin the enjoyment, which hints that the element of surprise is key for the pleasure when consuming these media. This is not so much the case for music. Music fans, when going to a concert, do not avoid listening to the set list beforehand, but rather the opposite, some fans tend to repeat a given album or playlist numerous times, in preparation for the incoming concert. In this context, fans are employing repetition as a means to maximize their enjoyment at the time of the event. In fact, they are making use of a well-documented relationship between the number of repetitions (or exposures) to the liking of a given piece of music, the so-called mere exposure effect~\cite{bornstein1989exposure}.

The mere exposure effect in music is a quite common phenomenon, where after listening to a song a couple of times in gatherings or supermarkets without clear attention, the individual starts to recognize it and then even to like it.  The effect, which has been long observed and studied by psychologists, states that the mere exposure of the individual to a given stimulus results in an increase of the positive affect (read liking, pleasantness) towards said stimulus~\cite{zajonc1968attitudinal}. Experiments on the effect show that people tend to prefer items more familiar to them, on account to the number of exposures, and is robust enough to be observed for different types of stimuli, be it images, words, simple chords to even more complex music~\cite{bornstein1989exposure}. Logically, however, while repetition can increase familiarity and in consequence the liking of a given stimulus, affect cannot increase monotonically with repetition, overexposure leads to satiation and boredom. The combination of the rising positive effect for liking with exposures coupled with the eventual satiation, result in the well-studied inverted-U curve, one of the most tested hypotheses for music preference~\cite{chmiel2017back}.


In recent years, music streaming platforms have revolutionized the way humans interact with music. Every day, millions of users  interact with a seemingly infinite music catalog. In this context, recommender systems help users navigate this catalog, directing them towards music they are likely to enjoy. Even if streaming platforms allow for music to be employed as never before, listening to music is still a human experience. The principles governing why we like a piece of music are still constricted to the functioning of the human mind. Ultimately, music recommender systems are merely exposing users to music. What this means is that the data corresponding to users interacting to (repetitive) recommendation is a bigger collection of data, on the mere exposure effect than previous generations of researchers on the subject could dream of. This allows for leveraging the vast body of knowledge advanced on the subject to interpret the data and develop models embedded with understanding rather than only driven by precision. 

In this paper, in the first part, we collect data on newly released song consumption from a major streaming platform. We show that the empirical probability of listening to these songs, which we employ as a proxy of interest, does evolve with the typical inverted-U shape of the mere exposure effect. Later on, we discuss how this behavior can be exploited for characterizing both music and users of music streaming platforms. We finish with a discussion on how future work shall advance on these preliminary findings and how they can be employed for developing new recommendation paradigms. 

\section{Related Work}
\label{sec:related work}
\textbf{Mere exposure effect}. Among the most influential work devoted to the mere exposure effect, stands out Bornstein's meta-analysis of over 20 years of studies dedicated to it~\cite{bornstein1989exposure}. Bornstein’s work elaborates on how different factors, such as stimulus type and complexity, presentation sequence and individual traits, such as age and personality, modulate the magnitude of the effect. Bornstein also identifies satiation/boredom as a limiting factor to the effect, resulting in an inverted-U shaped relation between exposures and liking: the first exposures result in a positively increase on the liking of a stimulus, as exposures increase, liking attains a peak point where further exposures will only decrease it. 
For accounting to this phenomenon, Bornstein then calls attention to the two factor model for stimuli preference proposed by Berlyne~\cite{berlyne1970novelty}. This model describes the mere exposure effect as a combination of: (1) an habituation factor, increasing affect towards the stimulus as it becomes more familiar and (2) a tedium factor resulting from over exposure. When combined, these two processes result in the inverted-U shape. Berlyne goes even beyond, by proposing a theory not only restricted to familiarity, where the hedonic value (read pleasure) of a stimulus varies, following the inverted-U shape in function of a number of ``collative variables'' such as novelty/familiarity and complexity~\cite{chmiel2017back}. 
More recently, Montoya et al~\cite{montoya2017re}, inspired by Bornstein, presented a novel meta-analysis, collecting  research articles on the 2 decades of the evidence and theoretical development on the exposure effect which follows Bornstein`s paper. The article explores a number of facets modulating the effect and challenges four theories explaining it (among which Berlyne’s two factor one). When data allowed for it, they estimated the growth curve for each investigation via a quadratic regression of the mean ratings (liking, pleasing, recognition, etc.) of the stimulus on the number of exposures and number of exposures squared. One of the key findings is that, across the 81 investigated articles, and over 268 analyzed curves, the mere exposure effect can be characterized by a positive slope and negative quadratic effect consistent with the inverted-U shaped curve. Interestingly enough, the obtained fit demonstrates a negative slope and positive quadratic term, indicating a U shaped curve, only for the auditory stimuli (which is the main target of this paper). 

Considering musical stimuli, in~\cite{chmiel2017back}, Chmiel and Schubert, based in Berlyne’s theory, investigate the pertinence of the inverted-U curve on a corpus of over 100 years of research involving music preference as functions of collative variables, in particular familiarity (often using exposures as a proxy) and complexity. They report that fifty out of the fifty seven considered studies report findings compatible with either the inverted-U model or a segmented version of it, where individuals find themselves to the left or right from the peak point. 


\textbf{Repeat consumption}. Benson et al.~\cite{benson2016modeling} define repetitive consumption as the process by which users discover new items, become enamored with it, consume it frequently and then transfer their attention to other alternatives. They study repetitive consumption in different domains, one of which, music streaming. Benson reports that very few items are consumed many times (heavy-tailed distribution), and that eventually, the gap between reconsumptions tends to grow bigger, a process interpreted as an increasing boredom.  They proposed a model that is able to capture both of these aspects. 
In~\cite{anderson2014dynamics}, Anderson also explores repetitive patterns of consumption in different domains, 
among which radio playlists. Based on their investigation on factors modulating repetitive consumption throughout all the datasets, the authors introduce a reconsumption model, based on recency, where 
users interact with new items for a period of interest, before growing tired and eventually discarding them. They later add a ``quality’’ factor (popular or high-quality items are more likely to be consumed) to their model, therefore predicting reconsumption based on both regency and quality. A key difference between Benson’s and Anderson’s and the model is that the latter is time-independent.

Other than statistical modeling, repetition consumption is an interesting topic for the development of recommender systems, as users tend to interact with the same item repeatedly in a wide set of domains~\cite{tsukuda2020explainable}. There are a number of studies that consider repetition or its modulating factors, such as boredom, when implementing new recommender frameworks, e.g.~\cite{hu2011nextone,rafailidis2015repeat,chmiel2018simple,ren2019repeatnet}. From a perspective perhaps closer to the aim of this paper, authors introduce in~\cite{reiter2021predicting} a psychological-informed approach for predicting music relistening. Inspired on psychology studies showing that music preference is biased by memory. The authors propose a recommendation model based on a proxy of human memory given by the cognitive architecture ACT-R~\cite{anderson2009can}. Among the different implemented components from ACT-R is the base level activation component, which is an activation energy that increases when an item is reconsumed and slowly decay over time, in a power-law fashion. This component, by itself allows for the accounting for either the frequency and recency of previously interacted songs and predict relistening behavior.

The body of research discussed above shows, either an exploration of the mere exposure effect, or the importance of accounting for repetitive behavior for consumption and recommendation. However, rarely both topics appear in the same work. Studies such as~\cite{chmiel2018simple, chmiel2018using} do indeed discuss how Berlyne's two factor model could be employed for generating music recommendation. However they are mostly concerned with avoiding over-exposure and lack empirical validation. In this article, we propose a new approach, bridging both subjects and employing the mere exposure effect as an investigating tool for leveraging information from repetitive behavior. As next sessions will develop, there are a number of factors that modulate the effect that are encoded in the daily user behavior on music streaming platforms, we believe that these patterns are of capital importance for the characterization of music and eventual development of new music recommendation paradigms.

\section{Data}
\subsection{Dataset}
To study the exposure effect in music consumption, we collected listening logs of users from XXXX\footnote{Name kept hidden for anonymous purposes}, one of the leading streaming platforms. However, if one is to observe the inverted-U shape characteristic of the mere exposure effect, a key factor of it is, as put by Berlyne, the habituation factor, i.e., getting used to or, in other words, ``learning'' the stimulus. Without this factor, we are left with the  tedium factor, decreasing interest over exposures. What this means is that of most importance here is the investigation of the first exposures of a user to a given song, i.e., the repetitive consumption of new and fresh items (at least to the user). In order to decrease the likelihood of a user already knowing a given track beforehand, we collected data of individuals interacting with a set of seven albums released on November 19, 2021 for the two month period following the release. The albums' genres vary from French Rap to R\&B and international pop music, and sum to 124 unique tracks. 

Another variable of interest is the number of exposures to be considered. In~\cite{bornstein1989exposure}, Bornstein points out that on the work investigated, the number of expositions on the examined work vary from 10 to 50, with the mean ceiling being 20.95 (SD = 32.28). Which, although relatively low, was sufficient for many researchers to find a decline in ratings. We therefore will consider only users who were in contact with one of the selected tracks for 40 exposures (in case of subsequent exposures after the first 40, the data was discarded). For users of a streaming service, 40 exposures in a period of 2 months is relatively high, and therefore we are mostly accounting for users that are fans of the considered artists. However, we argue that accounting for a longer period of time to collect the 40 exposures has some drawbacks, for instance, it increases the likelihood of a given user discovering these tracks through some other media, say social networks. Moreover, it allows for forgetting mechanisms to come into play, which might hinder the tedium factor, allowing for a slower decay of interest.  

The resulting dataset accounts for the data of 125K unique users who were exposed to  at least one of the 124 tracks 40 times. The dataset has the following form: every line corresponds to one exposure of a pair user-track with a corresponding id for both, the ordered number of exposure, the timestamp of when the exposure happened and the total listening time. We add a binary variable ``listened'' $l(s)$ for each line, indicating that the song $s$ was listened, $l(s) =  1$ if the listening time is superior or equal to 30s, and 0 otherwise (a threshold largely employed by the industry as the definition of one listening event for remuneration purposes). The intuition being that if the user is not interested on a given song, they will not listen to more than 30s of it, therefore we regard the listening of a song as an indication of the user's interest.  The resulting dataset has over 9M lines (note that some users were exposed to more than one of the 124 tracks for 40 times). `

\subsection{The Inverted-U curve in music discovery}
\begin{figure} 
    \centering
    \begin{minipage}{.3\linewidth}
        \centering
        \includegraphics[width=1\linewidth]{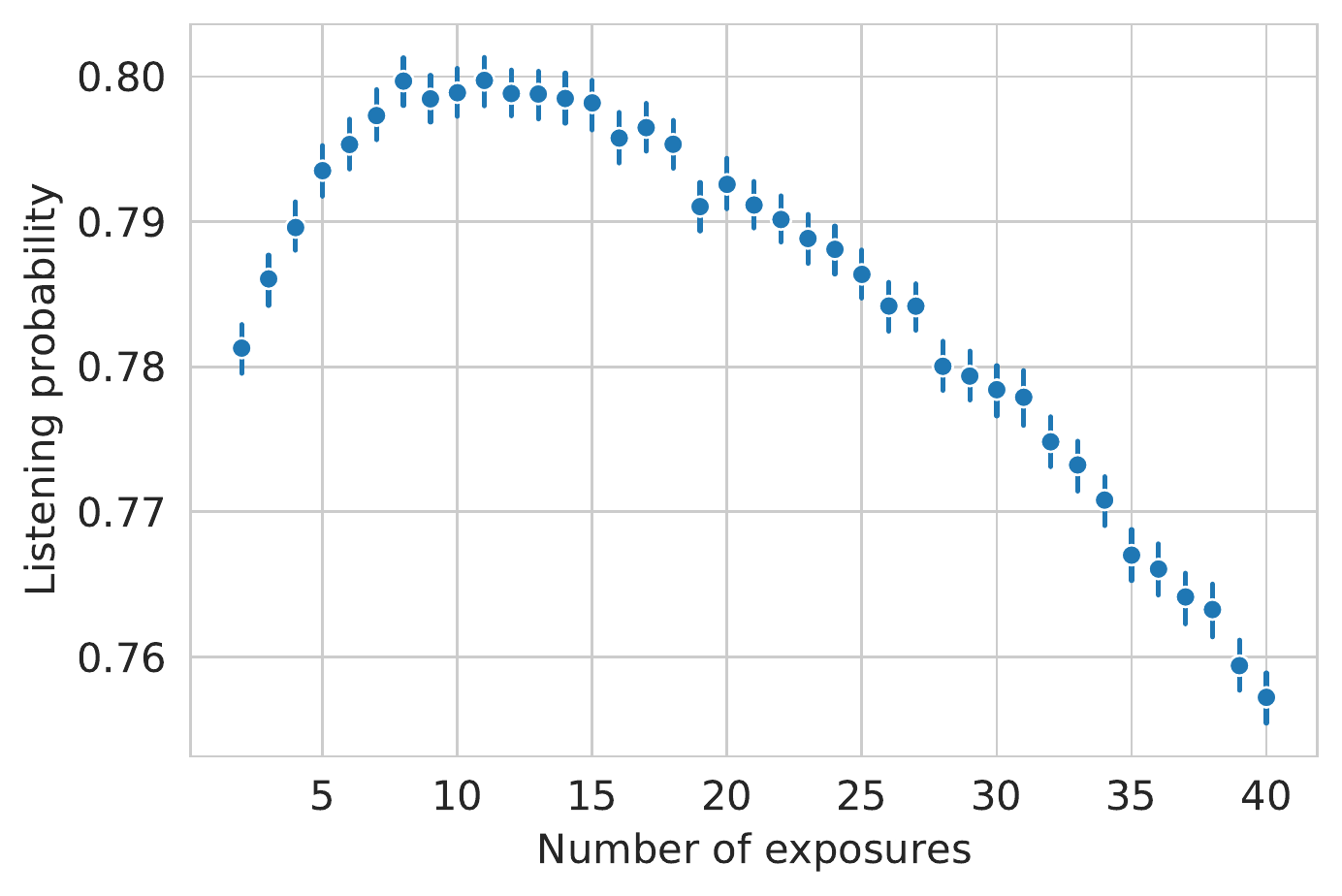}
    \end{minipage}%
    \begin{minipage}{0.3\linewidth}
        \centering
        \includegraphics[width=1\linewidth]{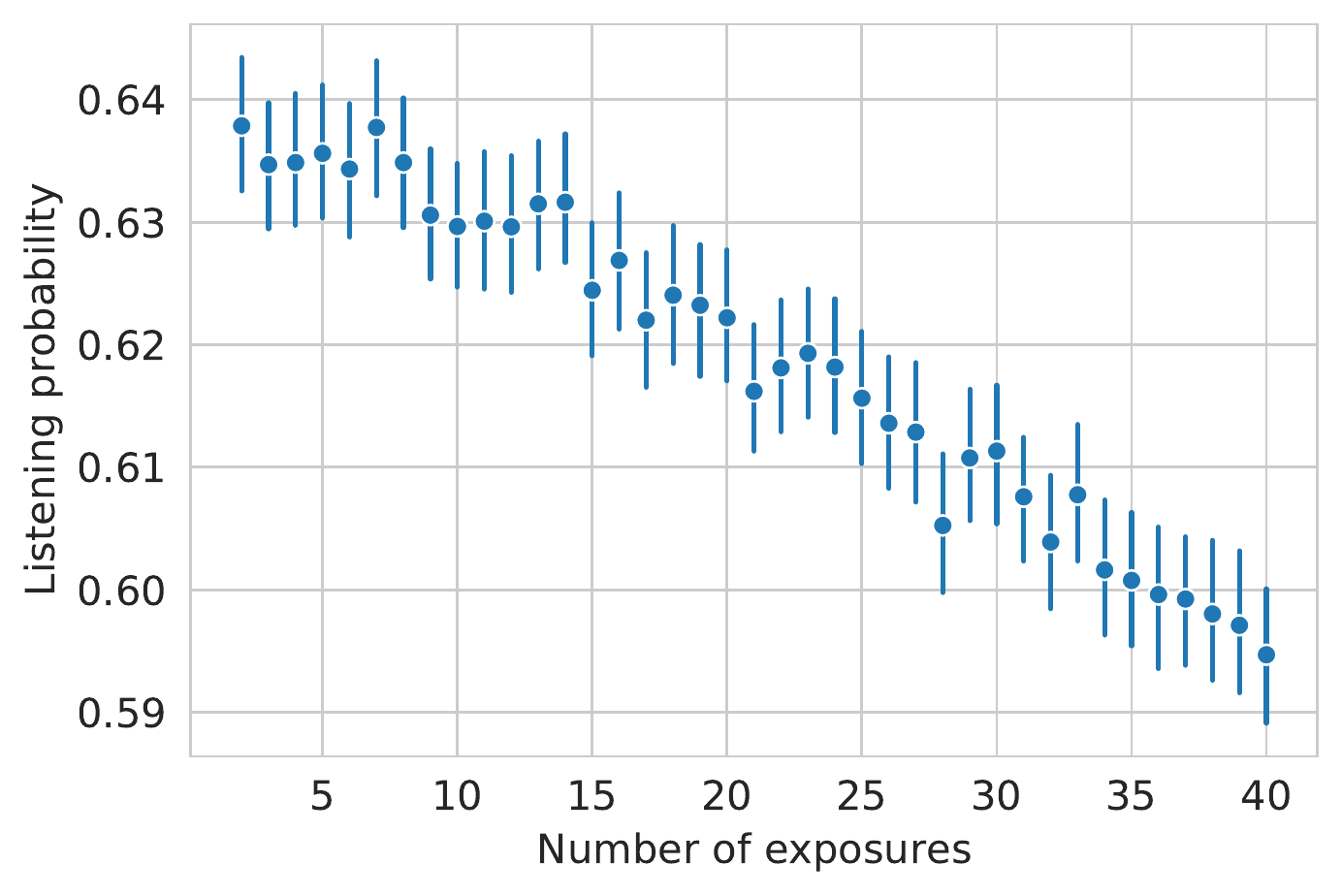}
    \end{minipage}
    \caption{Left: Aggregated listening probability with 95\% confidence intervals over the number of exposures from the consumption of new released albums. Right: Aggregated listening probability with 95\% confidence intervals over the number of exposures for famous (and therefore known) rock albums.}
    \label{fig:inverted-U-all}
\end{figure}

As mentioned above, when exposed to a new stimulus, interest increases with familiarity on the account of the number of exposures, attains a peak after which, new exposures will only increase satiation, resulting in the inverted-U shape. It is then plausible that users listening to new songs will increase their interest after a couple of exposures and then grow tired of it. We check to see whether we observe this behavior in the collected data by computing the aggregated empirical listening probability for all users at their $x$th exposure: $P(l(s) = 1|x)$. On the left side of Figure~\ref{fig:inverted-U-all} it depicts the evolution of the listening probability (represented by the mean value and the confidence intervals) over the number of exposures for all users and all songs in the collected dataset. The right side of Figure~\ref{fig:inverted-U-all} also depicts the evolution of the listening probability, but this time for 30 classic rock albums such as The Beatles' Let it Be, Pink Floyd's Dark Side of the Moon and Nirvana's Nevermind (selected to increase the likelihood of the users already knowing these songs, as they were all released at least 25 years ago). The listening probability of the right curve was computed from streaming data of 19K unique users (not necessarily the same as the previous dataset) streaming 360 songs for the same two months as the  dataset described above.  While for the classic albums there's a slow decay of the listening probability over exposures (satiation), for the new releases data, a trend is clearly observable, where the listening probability goes up after a couple of exposures, attains a peak point and then slowly decays with the subsequent exposures, very much in accord with the mere exposure effect.  Note that here (and in subsequent figures) we are accounting for the behavior after the first exposure ($x>1$) as we posit that a user can listen to a given song the first time out of curiosity (specially for the newly released albums), however, the second exposure is more representative of their actual interest. 


Montoya et al. in~\cite{montoya2017re}, employed a quadratic regression fit to evaluate the inverted-U shape (positive slope and negative quadratic term) on a number of studies where individual's preferences over the course of exposures are reported. Given that our assessed variable $l(s)$ is binary, we cannot do the same. Instead, to evaluate the shape, we fit a probit model over the consumption data. The probit model is a traditionally employed regression model in statistics, where the output variable is binary. Here, it is assumed that a binary variable $y$ is driven by a latent dependent variable $y^*$ which is a linear combination of some explanatory variables $X$: $P(y = 1|X) = \Phi(y^*)$, where $y^* = X^T \beta + \epsilon$, $\beta$ is a vector of estimable parameters, $\epsilon$ is the disturbance term, assumed to be normally distributed with mean 0 and variance 1 and $\Phi(z)$ is the cumulative normal distribution from $-\infty$ to $z$. 
Then, y is an indication if the latent $y^*$ is positive or not, $y = 1$, if $y^*>0$, 0 otherwise. The idea here is that the underlying decision process controlling a listening event has a hard threshold on the interest, but that the precise location of the threshold is subject to random Gaussian noise.

We set the exposition number $x$ as the linear term, a quadratic term $x^2$ and a constant $1$ as the explainable variables $X$ of the probit model. Therefore, much as~\citet{montoya2017re}, we are defining the latent variable $y^*$ as a quadratic function over the number of exposures, in this case $y^*$ is a proxy of the user interest and $y = l(s)$ is a noisy observation on it. The fit model is represented by the equation $y^* = 0.7952 + 0.0048x -0.0002x^2$, the coefficients were computed by maximum likelihood and have $ p< 0.001$. 
From the fit coefficients, the interest  peaks at 12 exposures, and although the quadratic function cannot completely model the data (the shape of the curve not being symmetrical), the obtained coefficients for the positive slope and negative quadratic term are both characteristic of the famous inverted-U. Note the high intercept close to 0.8, indicating that there users are mostly fans of the content (as they are the first ones to listen to the new albums for a high number of times). Fitting the same model for the data of on the right of Figure~\ref{fig:inverted-U-all}, result in a peak around $-32$ exposures, although this values does not make sense, it is to be expected, as most users already know these songs and therefore there's no habituating factor into play (rise of interest), only the tedium one.

\section{Music and Users Factors Modulating the Mere Exposure Effect}
Last section was concerned with finding whether data of users consuming new songs do indeed follow the characteristic inverted-U shape of the mere exposure effect. In this section we explore how this behavior can be employed for the characterization of both music and users. As mentioned beforehand, there are a number of factors modulating the magnitude of the mere exposure effect, citing a few from~\cite{bornstein1989exposure}: stimulus complexity, individual’s personality and age, and presentation sequence and duration. Here we posit that all these factors are encoded in the observed behavior. In fact, the resulting curves are the aggregation of individual pairs of user-song, all presenting different forms conditional on both the song’s and user’s characteristics. 

 \begin{figure}
    \centering
    \includegraphics[width=0.5\linewidth]{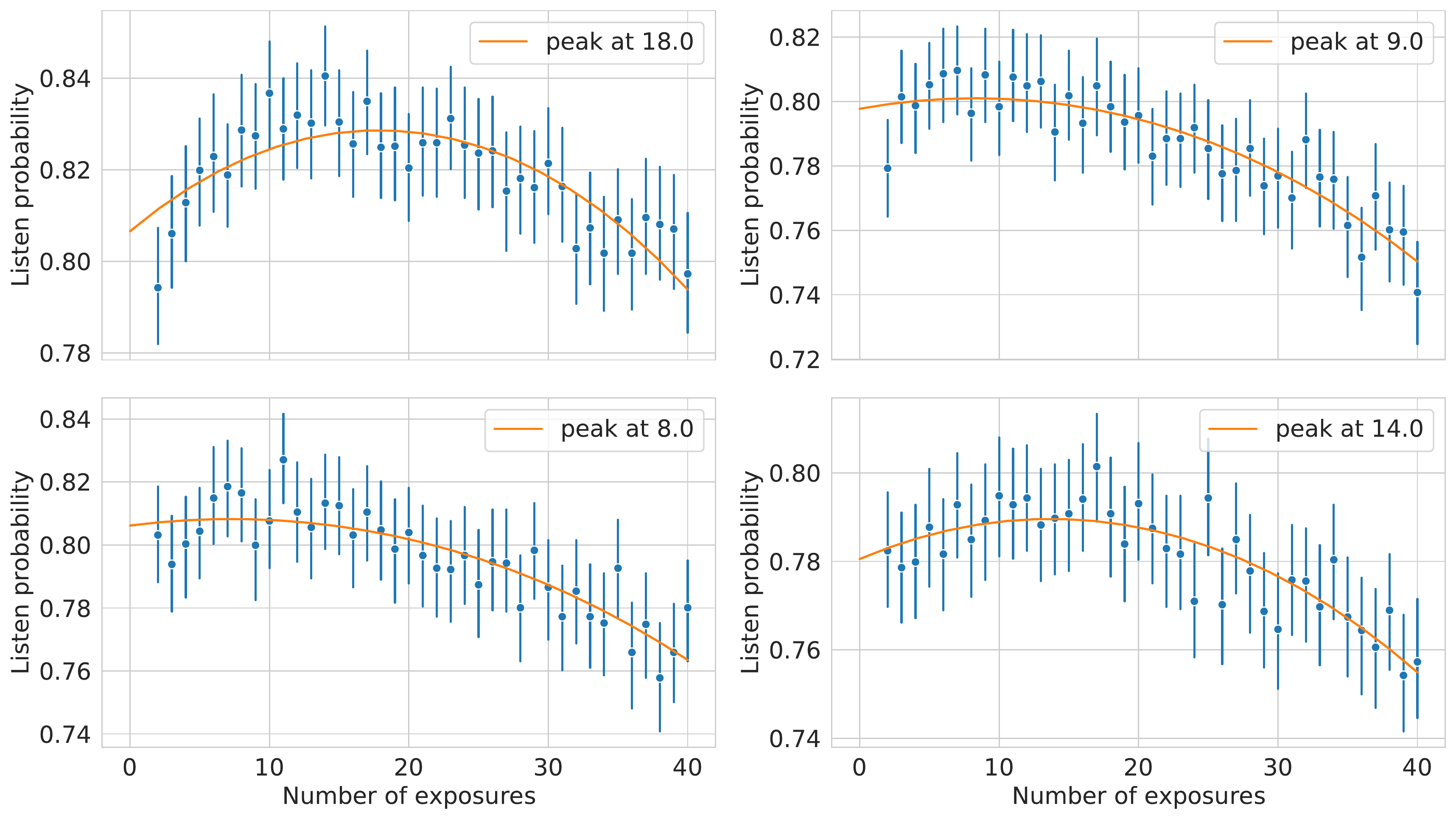}
    \caption{Aggregated listening probability over the number of exposures for four tracks with the estimated probability of listening per exposure when fitting the probit model.}
\label{fig:inverted-U}
\end{figure}

Figure~\ref{fig:inverted-U} depicts the aggregated listening probability, however, this time for four different tracks. Here we chose four tracks from the Album ``30'' of the English song writer Adele (one of the albums in our dataset), to increase the co-occurrence of the same users and diminish individual differences that might modulate the shape. We also included the fit $\Phi(y^*)$ obtained with the probit model, as described above, but for each individual track. Note that although the quadratic probit model fails to precisely model the data, we include it to obtain an estimation of the effect's peak. It is noticeable that, while they all have the characteristic inverted-U shape, they peak at a different number of exposures. This fact is rich with information. For example, a song whose peak is attained with little to no exposures is supposedly, by default, more familiar to these users (much as the right side of Figure~\ref{fig:inverted-U-all}), while a song that takes a higher number of exposures to attain the peak was, at first, a less familiar one (left of Figure~\ref{fig:inverted-U-all}). Although these songs are new (we're not considering ``Easy on me'', a track from the album previously released as a single) and from the same artist, they can be more or less similar to what the fans are used to. The number of exposures to attain the peak for a given group of users is, therefore, a proxy of these users' familiarity to that given song. 

 \begin{figure} [h]
    \centering
    \includegraphics[width=0.6\linewidth]{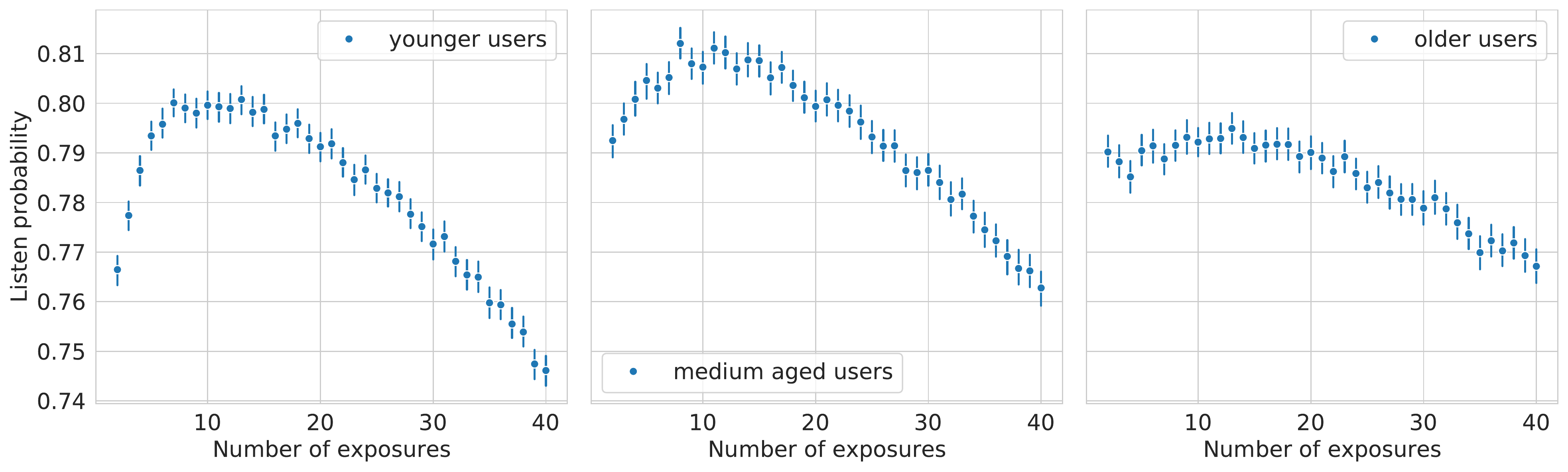}
    \caption{Aggregated listening probability over the number of exposures for users split by age.}
\label{fig:inverted-U-age}
\end{figure}

However, the inverted-U shape is not only driven by familiarity, individual traits are also a modulating factor of the effect. One of the individual traits that have an impact on the magnitude of the mere exposure effect, discussed by Bornstein in his meta analysis, is the individual's age. We split the users in our dataset into 3 classes with roughly the same number of individuals (filtering out users whose reported age was either missing or too far from the norm) , the classes are: younger users (43K users, age $\leq$ 21), medium aged users (32K users, 21 > age $\leq$ 27)  and older users (34K users, age > 27). Figure~\ref{fig:inverted-U-age} depicts the empirical probability aggregated by the users of each age category. We note clear differences between the dynamics in each class. Younger users tend to have a more expressive U shaped consumption, with a greater increase/decrease of interest than the other two. This behavior indicates that younger users, when discovering new music, take more risk, listening to songs they are not completely interested in at first (lower intercept than the other two groups), however their interest peaks after a couple listening events, later on decreasing faster than the other two classes. This more dynamic exploratory behavior of the youngsters measures up to previous research in music psychology where it is shown that music preferences become more stable, and less prone for exploration, as individuals age~\cite{knees2019user}.

The examples above serve as illustrations of the many factors both content and user-based, that shape the interest's evolution. An interesting factor not yet explored here is complexity. In~\cite{bornstein1989exposure, montoya2017re, szpunar2004liking}, complexity is seen as a modulating factor of the mere exposure effect, with higher stimuli complexity resulting in a more expressive effect. Although measuring music complexity is not straightforward, 
the dynamics of the mere exposure effect might be a way for assessing users' perceived complexity. Due to the vast available data of users discovering new songs daily in music streaming platforms, the dynamics of the interest curve for a group of users or songs can be then easily obtained and exploited. Understanding these dynamics allows for the characterization of users who are more interested in novelty (such as the young users in Figure~\ref{fig:inverted-U-age}), or songs seemingly more complex, taking more expositions for interest to peak. This allows for modulating recommendation based on both the user profile and the song's characteristics.  





\section{Conclusion}
\label{sec: conclusion}
In this paper we showed that when exposed to new songs repeatedly, users' interest evolves with the well-studied inverted-U shape of the mere exposure effect. This fact allows for the leveraging of the vast body of research on the subject to understand consuming data, characterize both music and users and develop new recommendation paradigms. Moreover, the mere exposure effect serves as a novel optic for evaluating music discovery. As we have shown, a single exposition is not enough for the user for the interest to peak, or for a song to be``learned''. Therefore, we believe that accounting for multiple repetitions is of most importance when new music is presented.

Future work should develop models to estimate the evolution of the interest curve for a given pair user-song based on the user's historical data and the history of other users consuming  said song. If shown to be effective, then we can estimate the number of expositions for the user's interest to attain the peak, which gives us both an estimation on how familiar the given track is for the user, but also the potential for repetitive recommendation. This allows for recommendations to be made, balancing familiar and less familiar songs, repetition and exploration, to help users discover music.


\bibliographystyle{ACM-Reference-Format}
\bibliography{repetition}


\begin{thebibliography}{17}


\ifx \showCODEN    \undefined \def \showCODEN     #1{\unskip}     \fi
\ifx \showDOI      \undefined \def \showDOI       #1{#1}\fi
\ifx \showISBNx    \undefined \def \showISBNx     #1{\unskip}     \fi
\ifx \showISBNxiii \undefined \def \showISBNxiii  #1{\unskip}     \fi
\ifx \showISSN     \undefined \def \showISSN      #1{\unskip}     \fi
\ifx \showLCCN     \undefined \def \showLCCN      #1{\unskip}     \fi
\ifx \shownote     \undefined \def \shownote      #1{#1}          \fi
\ifx \showarticletitle \undefined \def \showarticletitle #1{#1}   \fi
\ifx \showURL      \undefined \def \showURL       {\relax}        \fi
\providecommand\bibfield[2]{#2}
\providecommand\bibinfo[2]{#2}
\providecommand\natexlab[1]{#1}
\providecommand\showeprint[2][]{arXiv:#2}

\bibitem[\protect\citeauthoryear{Anderson, Kumar, Tomkins, and
  Vassilvitskii}{Anderson et~al\mbox{.}}{2014}]%
        {anderson2014dynamics}
\bibfield{author}{\bibinfo{person}{Ashton Anderson}, \bibinfo{person}{Ravi
  Kumar}, \bibinfo{person}{Andrew Tomkins}, {and} \bibinfo{person}{Sergei
  Vassilvitskii}.} \bibinfo{year}{2014}\natexlab{}.
\newblock \showarticletitle{The dynamics of repeat consumption}. In
  \bibinfo{booktitle}{\emph{Proceedings of the 23rd international conference on
  World wide web}}. \bibinfo{pages}{419--430}.
\newblock


\bibitem[\protect\citeauthoryear{Anderson}{Anderson}{2009}]%
        {anderson2009can}
\bibfield{author}{\bibinfo{person}{John~R Anderson}.}
  \bibinfo{year}{2009}\natexlab{}.
\newblock \bibinfo{booktitle}{\emph{How can the human mind occur in the
  physical universe?}}
\newblock \bibinfo{publisher}{Oxford University Press}.
\newblock


\bibitem[\protect\citeauthoryear{Benson, Kumar, and Tomkins}{Benson
  et~al\mbox{.}}{2016}]%
        {benson2016modeling}
\bibfield{author}{\bibinfo{person}{Austin~R Benson}, \bibinfo{person}{Ravi
  Kumar}, {and} \bibinfo{person}{Andrew Tomkins}.}
  \bibinfo{year}{2016}\natexlab{}.
\newblock \showarticletitle{Modeling user consumption sequences}. In
  \bibinfo{booktitle}{\emph{Proceedings of the 25th International Conference on
  World Wide Web}}. \bibinfo{pages}{519--529}.
\newblock


\bibitem[\protect\citeauthoryear{Berlyne}{Berlyne}{1970}]%
        {berlyne1970novelty}
\bibfield{author}{\bibinfo{person}{Daniel~E Berlyne}.}
  \bibinfo{year}{1970}\natexlab{}.
\newblock \showarticletitle{Novelty, complexity, and hedonic value}.
\newblock \bibinfo{journal}{\emph{Perception \& psychophysics}}
  \bibinfo{volume}{8}, \bibinfo{number}{5} (\bibinfo{year}{1970}),
  \bibinfo{pages}{279--286}.
\newblock


\bibitem[\protect\citeauthoryear{Bornstein}{Bornstein}{1989}]%
        {bornstein1989exposure}
\bibfield{author}{\bibinfo{person}{Robert~F Bornstein}.}
  \bibinfo{year}{1989}\natexlab{}.
\newblock \showarticletitle{Exposure and affect: overview and meta-analysis of
  research, 1968--1987.}
\newblock \bibinfo{journal}{\emph{Psychological bulletin}}
  \bibinfo{volume}{106}, \bibinfo{number}{2} (\bibinfo{year}{1989}),
  \bibinfo{pages}{265}.
\newblock


\bibitem[\protect\citeauthoryear{Chmiel and Schubert}{Chmiel and
  Schubert}{2017}]%
        {chmiel2017back}
\bibfield{author}{\bibinfo{person}{Anthony Chmiel} {and} \bibinfo{person}{Emery
  Schubert}.} \bibinfo{year}{2017}\natexlab{}.
\newblock \showarticletitle{Back to the inverted-U for music preference: A
  review of the literature}.
\newblock \bibinfo{journal}{\emph{Psychology of Music}} \bibinfo{volume}{45},
  \bibinfo{number}{6} (\bibinfo{year}{2017}), \bibinfo{pages}{886--909}.
\newblock


\bibitem[\protect\citeauthoryear{Chmiel and Schubert}{Chmiel and
  Schubert}{2018a}]%
        {chmiel2018simple}
\bibfield{author}{\bibinfo{person}{Anthony Chmiel} {and} \bibinfo{person}{Emery
  Schubert}.} \bibinfo{year}{2018}\natexlab{a}.
\newblock \showarticletitle{A simple algorithm for music recommendation, built
  on established psychological principles}. In \bibinfo{booktitle}{\emph{New
  music concepts: 5th International conference, ICNMC}}.
  \bibinfo{pages}{155--160}.
\newblock


\bibitem[\protect\citeauthoryear{Chmiel and Schubert}{Chmiel and
  Schubert}{2018b}]%
        {chmiel2018using}
\bibfield{author}{\bibinfo{person}{Anthony Chmiel} {and} \bibinfo{person}{Emery
  Schubert}.} \bibinfo{year}{2018}\natexlab{b}.
\newblock \showarticletitle{Using psychological principles of memory storage
  and preference to improve music recommendation systems}.
\newblock \bibinfo{journal}{\emph{Leonardo Music Journal}}
  \bibinfo{volume}{28} (\bibinfo{year}{2018}), \bibinfo{pages}{77--81}.
\newblock


\bibitem[\protect\citeauthoryear{Hu and Ogihara}{Hu and Ogihara}{2011}]%
        {hu2011nextone}
\bibfield{author}{\bibinfo{person}{Yajie Hu} {and} \bibinfo{person}{Mitsunori
  Ogihara}.} \bibinfo{year}{2011}\natexlab{}.
\newblock \showarticletitle{NextOne Player: A Music Recommendation System Based
  on User Behavior.}. In \bibinfo{booktitle}{\emph{ISMIR}},
  Vol.~\bibinfo{volume}{11}. \bibinfo{pages}{103--108}.
\newblock


\bibitem[\protect\citeauthoryear{Knees, Schedl, Ferwerda, and Laplante}{Knees
  et~al\mbox{.}}{2019}]%
        {knees2019user}
\bibfield{author}{\bibinfo{person}{Peter Knees}, \bibinfo{person}{Markus
  Schedl}, \bibinfo{person}{Bruce Ferwerda}, {and} \bibinfo{person}{Audrey
  Laplante}.} \bibinfo{year}{2019}\natexlab{}.
\newblock \showarticletitle{User awareness in music recommender systems}.
\newblock \bibinfo{journal}{\emph{Personalized human-computer interaction}}
  (\bibinfo{year}{2019}), \bibinfo{pages}{223--252}.
\newblock


\bibitem[\protect\citeauthoryear{Montoya, Horton, Vevea, Citkowicz, and
  Lauber}{Montoya et~al\mbox{.}}{2017}]%
        {montoya2017re}
\bibfield{author}{\bibinfo{person}{R~Matthew Montoya},
  \bibinfo{person}{Robert~S Horton}, \bibinfo{person}{Jack~L Vevea},
  \bibinfo{person}{Martyna Citkowicz}, {and} \bibinfo{person}{Elissa~A
  Lauber}.} \bibinfo{year}{2017}\natexlab{}.
\newblock \showarticletitle{A re-examination of the mere exposure effect: The
  influence of repeated exposure on recognition, familiarity, and liking.}
\newblock \bibinfo{journal}{\emph{Psychological bulletin}}
  \bibinfo{volume}{143}, \bibinfo{number}{5} (\bibinfo{year}{2017}),
  \bibinfo{pages}{459}.
\newblock


\bibitem[\protect\citeauthoryear{Rafailidis and Nanopoulos}{Rafailidis and
  Nanopoulos}{2015}]%
        {rafailidis2015repeat}
\bibfield{author}{\bibinfo{person}{Dimitrios Rafailidis} {and}
  \bibinfo{person}{Alexandros Nanopoulos}.} \bibinfo{year}{2015}\natexlab{}.
\newblock \showarticletitle{Repeat consumption recommendation based on users
  preference dynamics and side information}. In
  \bibinfo{booktitle}{\emph{Proceedings of the 24th International Conference on
  World Wide Web}}. \bibinfo{pages}{99--100}.
\newblock


\bibitem[\protect\citeauthoryear{Reiter-Haas, Parada-Cabaleiro, Schedl,
  Motamedi, Tkalcic, and Lex}{Reiter-Haas et~al\mbox{.}}{2021}]%
        {reiter2021predicting}
\bibfield{author}{\bibinfo{person}{Markus Reiter-Haas}, \bibinfo{person}{Emilia
  Parada-Cabaleiro}, \bibinfo{person}{Markus Schedl}, \bibinfo{person}{Elham
  Motamedi}, \bibinfo{person}{Marko Tkalcic}, {and} \bibinfo{person}{Elisabeth
  Lex}.} \bibinfo{year}{2021}\natexlab{}.
\newblock \showarticletitle{Predicting music relistening behavior using the
  ACT-R framework}. In \bibinfo{booktitle}{\emph{Fifteenth ACM Conference on
  Recommender Systems}}. \bibinfo{pages}{702--707}.
\newblock


\bibitem[\protect\citeauthoryear{Ren, Chen, Li, Ren, Ma, and De~Rijke}{Ren
  et~al\mbox{.}}{2019}]%
        {ren2019repeatnet}
\bibfield{author}{\bibinfo{person}{Pengjie Ren}, \bibinfo{person}{Zhumin Chen},
  \bibinfo{person}{Jing Li}, \bibinfo{person}{Zhaochun Ren},
  \bibinfo{person}{Jun Ma}, {and} \bibinfo{person}{Maarten De~Rijke}.}
  \bibinfo{year}{2019}\natexlab{}.
\newblock \showarticletitle{Repeatnet: A repeat aware neural recommendation
  machine for session-based recommendation}. In
  \bibinfo{booktitle}{\emph{Proceedings of the AAAI Conference on Artificial
  Intelligence}}, Vol.~\bibinfo{volume}{33}. \bibinfo{pages}{4806--4813}.
\newblock


\bibitem[\protect\citeauthoryear{Szpunar, Schellenberg, and Pliner}{Szpunar
  et~al\mbox{.}}{2004}]%
        {szpunar2004liking}
\bibfield{author}{\bibinfo{person}{Karl~K Szpunar}, \bibinfo{person}{E~Glenn
  Schellenberg}, {and} \bibinfo{person}{Patricia Pliner}.}
  \bibinfo{year}{2004}\natexlab{}.
\newblock \showarticletitle{Liking and memory for musical stimuli as a function
  of exposure.}
\newblock \bibinfo{journal}{\emph{Journal of Experimental Psychology: Learning,
  Memory, and Cognition}} \bibinfo{volume}{30}, \bibinfo{number}{2}
  (\bibinfo{year}{2004}), \bibinfo{pages}{370}.
\newblock


\bibitem[\protect\citeauthoryear{Tsukuda and Goto}{Tsukuda and Goto}{2020}]%
        {tsukuda2020explainable}
\bibfield{author}{\bibinfo{person}{Kosetsu Tsukuda} {and}
  \bibinfo{person}{Masataka Goto}.} \bibinfo{year}{2020}\natexlab{}.
\newblock \showarticletitle{Explainable recommendation for repeat consumption}.
  In \bibinfo{booktitle}{\emph{Fourteenth ACM Conference on Recommender
  Systems}}. \bibinfo{pages}{462--467}.
\newblock


\bibitem[\protect\citeauthoryear{Zajonc}{Zajonc}{1968}]%
        {zajonc1968attitudinal}
\bibfield{author}{\bibinfo{person}{Robert~B Zajonc}.}
  \bibinfo{year}{1968}\natexlab{}.
\newblock \showarticletitle{Attitudinal effects of mere exposure}.
\newblock \bibinfo{journal}{\emph{Journal of personality and social
  psychology}}  \bibinfo{volume}{9} (\bibinfo{year}{1968}),
  \bibinfo{pages}{1--27}.
\newblock


\end{thebibliography}
\end{document}